\documentclass[prb,showpacs,twocolumn,floatfix]{revtex4}
\usepackage{graphicx}
\begin{document}
\def\rhov{{\mbox{\boldmath{$\rho$}}}}
\def\tauv{{\mbox{\boldmath{$\tau$}}}}
\def\Lambdav{{\mbox{\boldmath{$\Lambda$}}}}
\def\sigmav{{\mbox{\boldmath{$\sigma$}}}}
\def\xiv{{\mbox{\boldmath{$\xi$}}}}
\def\chiv{{\mbox{\boldmath{$\chi$}}}}
\def\oh{{\scriptsize 1 \over \scriptsize 2}}
\def\ot{{\scriptsize 1 \over \scriptsize 3}}
\def\of{{\scriptsize 1 \over \scriptsize 4}}
\def\tf{{\scriptsize 3 \over \scriptsize 4}}

\noindent
{\bf Comment on ``Multiferroicity Induced by Dislocated Spin-Density
Waves"}

A recent Letter by Betouras {\it et al.} [\onlinecite{PRL}] (BGB)
proposed a phenomenological magnetoelectric (ME) interaction to 
explain the spontaneous polarization
${\bf P}$ in a class of materials of which YMn$_2$O$_5$ (YMO)
was taken to be the best exemplar.  They claimed 1) its behavior
agreed perfectly with their theory and 2) previous theories failed
in this regard.  Here I show that these
claims are incorrect and that their interaction is
unimportant in most systems.

Figure 1 shows {\bf P} measured by Kagomiya {\it et al.}
[\onlinecite{KAGOM}] compared to the microscopic theory of Chapon
{\it et al.}  [\onlinecite{CHAPON}]. The ME interaction proposed by BGB
gives {\bf P}=0 in the low-temperature ($T<20$K) incommensurate (IC)
phase and, as I will discuss, nearly zero in the high-temperature ($T>20$K)
commensurate (CM) phase.  Below I show that the Landau
approach [\onlinecite{US}] fits the data qualitatively.

\vspace{0.11 in}
\noindent
\begin{minipage} {3.9 cm}
\noindent
{\bf Fig. 1.}
(Based on Ref. \onlinecite{CHAPON}.)
${\bf P}$ vs $T$ for YMO.  Filled circles:
theory of Ref. \onlinecite{CHAPON}.  Thin line:
data of Kagomiya {\it et al.}[\onlinecite{KAGOM}]
The dashed line indicates the IC-CM phase boundary.
\end{minipage}
\hspace{0.15 in}
\begin{minipage} {3.8 cm}
\begin{center}
\includegraphics[width=3.8cm]{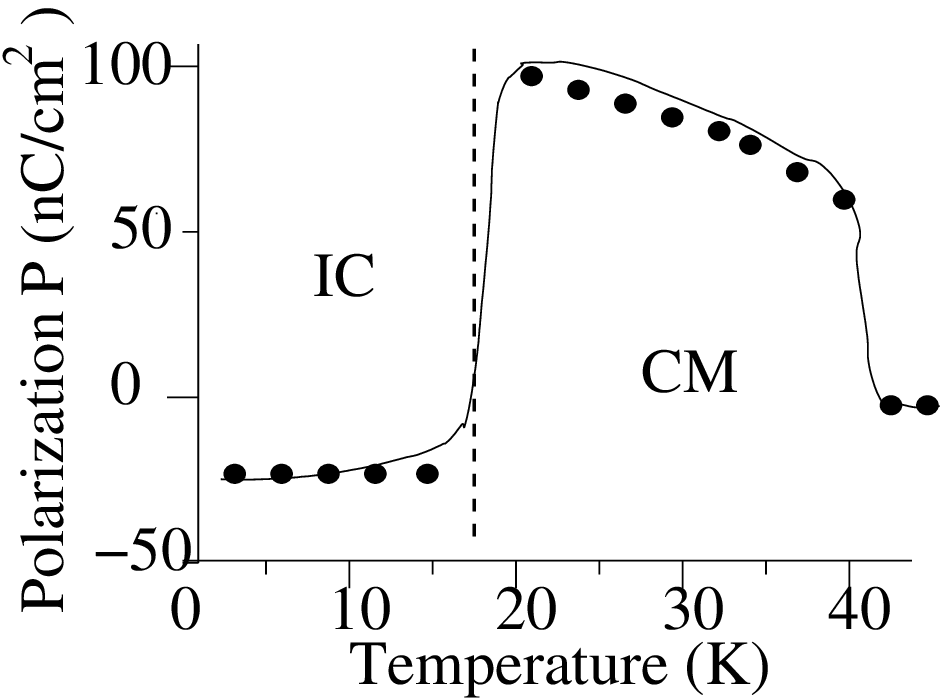}
\end{center}
\end{minipage}

\vspace{0.11 in}
The results of Ref. 1 follow from a gradient expansion
of the free energy to mimic the variation of magnetization
on an atomic length scale.  However,
the most successful long-wavelength treatment
of antiferromagnets [\onlinecite{HH}] utilizes the symmetry of
both the staggered and direct magnetization.  The generalization
to the multisublattice case of YMO is to consider all the
symmetry elements of the unit cell as done in Refs. 
\onlinecite{US}-\onlinecite{ABH}.  There one
introduced order parameters $\sigma_n$ of known symmetry to
describe the magnetism of the unit cell and for YMO
the trilinear ME coupling is [\onlinecite{ABH}]
$V = a P [|\sigma_1|^2 - |\sigma_2|^2]$, which gives
$P = {\rm const} \times [|\sigma_2|^2 - |\sigma_1|^2]$.  
I will call this the "standard" model and it was used
[\onlinecite{ABH}]
to successfully interpret the polarization of the CM
phase of YMO.  I show here that
this formulation can also explain the smaller nonzero polarization
in the IC phase which contradicts Ref. \onlinecite{PRL}.
For this purpose we need to extract the order parameters for YMO
since they were not given in Ref.  \onlinecite {CHAPON}. 
The symmetry of the spin pattern shown in their Fig. 2a
[\onlinecite{TOP}] for the CM phase indicates
that the order parameter called
$\sigma_1$ in Ref. \onlinecite{ABH} is zero. [\onlinecite{ABH}] To
normalize the results, I define $\sigma_2=1$.  Although
the size of the moments does not change much at the CM-IC 
transition, the orientations of the moments do change and
this change induces a significant redistribution of the order
parameters.  To see this consider the relation in their Fig. 2a
between the Mn$^{3+}$ spins at (0.09,-0.15,0.5) and at (0.41,0.35,0.5).
In the CM phase their $x$-components are the same but their
$y$-components change sign.  In the IC phase both $x$ and $y$-
components change sign. This change in spin orientation
induces significant changes in the order parameters.
I found the order parameters of the $T=1.9$K structure to be
such that $|\sigma_2|^2 - |\sigma_1|^2$
lies between -0.25 and +0.25. Even with this uncertainty
it is clear that the change in the magnetic structure
can explain the precipitate decrease in P at the CM to IC
phase boundary.

Since the microscopic model of Ref. \onlinecite{CHAPON} is so
satisfactory, one can ask whether it is consistent with either or both
of the competing phenomenological theories.  To answer this question
I write the ME interaction of BGB in the language of order
parameters effectively as
\begin{eqnarray}
V &=&{\rm const} \times 
P \sigma(q)^n 
\delta (n {\bf q}, {\bf G}) \ \ + \ {\rm c. \ c.} \ ,
\end{eqnarray}
where the $\delta$ function conserves momentum to within a
nonzero reciprocal lattice vector ${\bf G}$ (so that 
$n{\bf q}={\bf G}$).  For YMO, $n=4$ in Eq. (1), since
one of the magnetic unit cell axes is four times that of the 
paramagnetic unit cell.  The theory of Chapon {\it et al.} is
quadratic in the spin operators.  It can not give rise to
$4{\bf q}={\bf G}$, because this would require
either a four-spin ME interaction or treating the standard
two-spin ME interaction in second-order perturbation theory.
However, this latter possibility is quite
remote because the ME interaction is already quite weak in
systems where the ME interaction acts in first order perturbation
theory.  So the Chapon {\it et al.} model, which fits the
data nicely, is not a microscopic representation of the
ME interaction of BGB.  But since the model of Ref. 2 gives
the polarization as a quadratic form in the spin operators, it
is the microscopic analog of the "standard" model,
which I therefore claim does represent the data well.
Moreover, rather than happening in a wide class of
materials, the BGB interaction might
be important, but only when the magnetic unit
cell involves doubling of the paramagnetic unit cell.

In summary: the existing "standard" model explains the
data of Fig. 1 qualitatively, whereas that of BGB does not
and is unlikely to be important for most systems.

I thank A. P. Ramirez and T. Yildirim for suggestions.

\noindent
A. B. Harris

\noindent
Department of Physics and Astronomy

\noindent
University of Pennsylvania

\noindent
Philadelphia, PA 19104


\begin{thebibliography} {99}
\bibitem{PRL}
J. J. Betouras, G. Giovannetti, and J. van den Brink, Phys. Rev. Lett.
{\bf 98}, 257602 (2007).
\bibitem{CHAPON}
L. C. Chapon {\it et al.}, Phys. Rev. Lett. {\bf 96}, 097601 (2006).
\bibitem{KAGOM}
I. Kagomiya {\it et al.}, Ferroelectrics {\bf 286}, 167 (2003).
\bibitem{HH}
B. I. Halperin and P. C. Hohenberg, Phys. Rev. {\bf 188}, 898 (1969).
\bibitem{US}
G. Lawes {\it et al.}, Phys. Rev. Lett. {\bf 95}, 087205 (2005).
\bibitem{TMO}
M. Kenzelmann {\it et al.}, Phys. Rev. Lett. {\bf 95}, 087206 (2005).
\bibitem{PRB}
M. Kenzelmann {\it et al.}, Phys. Rev. B {\bf 74}, 014429 (2006). 
\bibitem{JAP}
A. B. Harris, J. Appl. Phys. {\bf 99}, 08E303 (2006).
\bibitem{ABH}
A. B. Harris, cond-mat/0610241.
\bibitem{TOP}
The top (bottom) panel should be labeled 24.7K (1.9K).
\end{thebibliography}
\end{document}